\documentclass[final]{aipproc}
\layoutstyle{8x11double}

\begin{document}

\title{Silicon-based spin quantum computation and the shallow donor exchange gate}

\author{Belita Koiller}{address={Instituto de F\'\i sica, Universidade Federal do Rio de Janeiro, Cx.  P. 68528, 21941-972 Rio de Janeiro, Brazil}}

\author{R. B. Capaz}{address={Instituto de F\'\i sica, Universidade Federal do Rio de Janeiro, Cx.  P. 68528, 21941-972 Rio de Janeiro, Brazil}}

\author{X. Hu}{address={Department of Physics, University at Buffalo, the State
University of New York, Buffalo, NY 14260-1500}}

\author{S. Das Sarma}{address={Condensed Matter Theory Center, Department of Physics,
University of Maryland, MD 20742-4111}}

\begin{abstract}
Proposed silicon-based quantum-computer architectures have attracted
attention because of their promise for scalability and their potential
for synergetically utilizing the available resources associated with the
existing Si technology infrastructure.
Electronic and nuclear spins of shallow donors (e.g. phosphorus) in
Si are ideally suited candidates for qubits in such proposals, where shallow
donor exchange gates are frequently invoked to perform two-qubit operations.
An important potential problem in this context is that intervalley
interference originating from the degeneracy in the Si conduction-band edge
causes fast oscillations in donor exchange coupling, which imposes
significant constraints on the Si quantum-computer architecture.
We discuss the theoretical origin of such oscillations. 
Considering two substitutional donors in Si, we present a systematic 
statistical study of the correlation between relative position distributions and the 
resulting exchange distributions. 
\end{abstract}

\maketitle


\section{Introduction}

As semiconductor devices decrease in size, their physical properties 
tend to become increasingly sensitive to the actual configuration of 
dopant substitutional impurities \cite{voyles}.  
A striking example is the proposal of donor-based
silicon quantum computer (QC) by Kane \cite{Kane}, in which the monovalent
$^{31}$P impurities in Si are the fundamental quantum bits (qubits).  This
intriguing proposal has created considerable recent interest in revisiting
all aspects of the donor impurity problem in silicon, 
particularly in the Si:$^{31}$P system.

Two-qubit operations for the donor-based Si QC architecture, 
which are required for a universal QC, involve precise control 
over electron-electron exchange\cite{Kane,
Vrijen,HD} and electron-nucleus hyperfine interactions (for nuclear spin qubits).  
Such control can presumably be achieved by fabrication of donor arrays with
accurate positioning and surface gates whose potential can be precisely
controlled \cite{Obrien,encapsulation,implant,schenkel03}.  
However, electron exchange in bulk silicon has spatial
oscillations\cite{Andres} on the atomic scale due to the valley interference
arising from
the particular six-fold degeneracy of the bulk Si conduction band.  These
oscillations place heavy burdens on device fabrication and coherent
control \cite{KHD1},
because of the very high accuracy requirement for placing each donor inside
the Si unit cell, and/or for controlling the external gate voltages.  

The potentially severe consequences of these problems for exchange-based Si QC
architecture motivated us and other researchers to perform further theoretical
studies, going beyond some of the simplifying approximations in the formalism
adopted in Ref.~\cite{KHD1}, and incorporating perturbation effects due
to applied strain\cite{KHD2} or gate fields \cite{wellard03}.  These
studies, performed within the standard Heitler-London (HL)
formalism \cite{slater}, essentially reconfirm the originally reported
difficulties regarding the sensitivity of the electron exchange coupling to
donor positioning, indicating that these may not be completely overcome by
applying strain or electric fields. 
The sensitivity of the calculated exchange coupling to donor relative 
position originates from interference between the plane-wave parts 
of the six degenerate Bloch states associated 
with the Si conduction-band minima. More recently \cite{Koiller2004} we have 
assessed the robustness of the HL approximation for the two-electron 
donor-pair states by relaxing the phase pinning at donor sites, 
which could in principle eliminate the oscillatory exchange behavior.
Within this more general theoretical scheme, the {\em floating-phase} HL
approach, 
our main conclusion is that, for all practical purposes, the
previously adopted HL wavefunctions are robust, and the
oscillatory behavior obtained in Refs.~\cite{KHD1,KHD2,wellard03}
cannot be taken as an artifact.

\section {Single donor}

We describe the single donor electron ground state using effective mass
theory.  The
bound donor electron Hamiltonian for an impurity at site ${\bf R}_0$ is
written as
\begin{equation}
{\cal H}_0={\cal H}_{SV}+{\cal H}_{VO} \,.
\label{eq:h0}
\end{equation}
The first term, ${\cal H}_{SV}$, is the single-valley Kohn-Luttinger
Hamiltonian \cite{Kohn}, 
which includes the single particle kinetic energy, the Si periodic potential,
and the impurity screened Coulomb perturbation potential 
\begin{equation}
V({\bf r})=-\frac{e^2}{\epsilon|{\bf r}-{\bf R}_0|} \,. 
\label{eq:coul}
\end{equation}
For shallow donors in Si, we use the static dielectric constant $\epsilon =
12.1$.  The second term of Eq.~(\ref{eq:h0}), ${\cal H}_{VO}$, represents the 
inter-valley scattering effects due to the presence of the impurity which 
breaks the bulk translational symmetry. 

\begin{figure}
\resizebox{75mm}{!}{\includegraphics[height=.3\textheight]{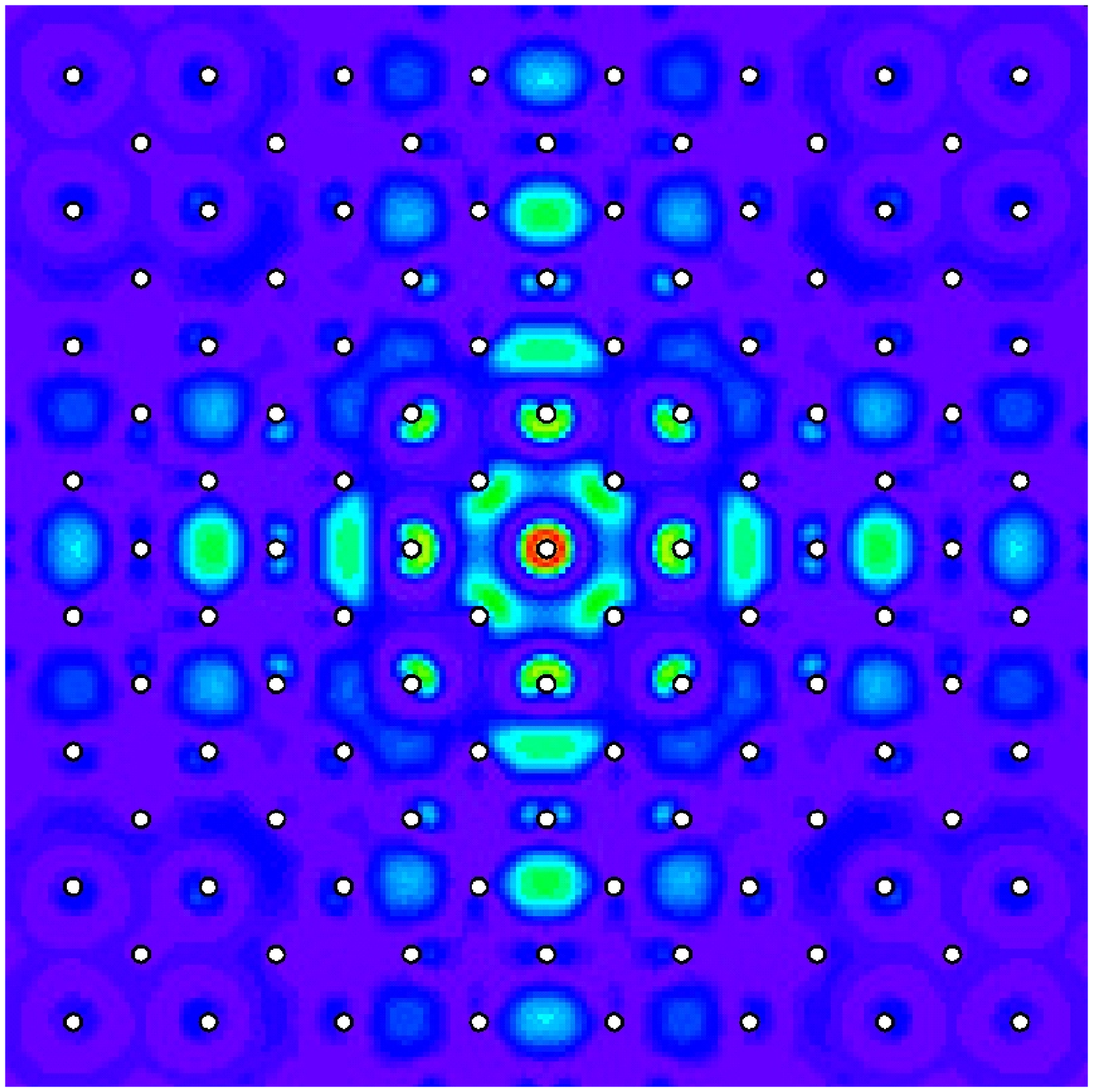}}
\caption{\label{charge}Electron probability density on the (001) plane of bulk Si 
for the ground state of a donor in Si within the envelope function approximation.
The white dots give the in-plane atomic sites.
The color scheme runs from red (high density) to purple (low density).
The highest density is at the central site, corresponding to the impurity site.}
\end{figure}

The electron eigenfunctions are written on the basis of the six unperturbed Si
band edge Bloch states $\phi_\mu = u_\mu(\bf r) e^{i {\bf k}_{\mu}\cdot {\bf
r}}$ [recall that the conduction band of bulk Si has six degenerate minima
$(\mu=1,\ldots,6)$, located along the $\Gamma-$X axes of the Brillouin zone
at $|{\bf k}_\mu|\sim 0.85(2\pi/{\rm a})$ from the $\Gamma$ point]:
\begin{equation}
\psi_{{\bf R}_0} ({\bf r}) = \frac{1}{\sqrt{6}}\sum_{\mu = 1}^6  F_{\mu}({\bf
r}-
{\bf R}_0) u_\mu({\bf r}) e^{i {\bf k}_{\mu}\cdot ({\bf r}-{\bf R}_0)}\,.
\label{eq:sim}
\end{equation}
The phases of the plane-wave part of all band edge Bloch states are naturally
chosen to be pinned at ${\bf R}_0$:  
In this way the charge density at the donor site
[where the donor perturbation potential Eq.~(\ref{eq:coul}) is more
attractive] is maximum, thus minimizing the energy for $\psi_{{\bf R}_0}({\bf
r})$.

In Eq.~(\ref{eq:sim}), $F_{\mu}({\bf r}-{\bf R}_0)$ are envelope functions 
centered at ${\bf R}_0$, for which we adopt the anisotropic Kohn-Luttinger
form, e.g., for $\mu = z$, 
$F_{z}({\bf r}) = \exp\{-[(x^2+y^2)/a^2 + z^2/b^2]^{1/2}\}/\sqrt{\pi a^2 b}$.
The effective Bohr radii $a$ and $b$ are variational parameters chosen to
minimize $E_{SV} = \langle\psi_{{\bf R}_0}| {\cal H}_{SV} |\psi_{{\bf
R}_0}\rangle$, leading to $a=25$ \AA, $b=14$ \AA~ and $E_{SV} \sim -30$ meV
when recently measured effective mass values are used in the
minimization \cite{KHD1}.
The periodic part of each Bloch
function is pinned to the lattice, independent of the donor site.  

The ${\cal H}_{SV}$ ground state is six-fold degenerate.  This degeneracy is
lifted by the valley-orbit interactions \cite{Pantelides}, which  
are included here in ${\cal H}_{VO}$, leading to the nondegenerate 
($A_1$-symmetry) ground state in (\ref{eq:sim}).
Fig.~\ref{charge} gives the charge density $|\psi_{{\bf R}_0} ({\bf r})|^2$ for this 
state, where the periodic part of the conduction band edge Bloch functions were obtained from 
{\it ab-initio} calculations, as described in Ref.~\cite{Koiller2004}. 
The impurity site ${\bf R}_0$, corresponding to the higher charge density, 
is at the center of the frame. It is interesting that, except for this central site, 
regions of high charge concentration and atomic sites do not necessarily coincide, 
because the charge distribution periodicity imposed by the plane-wave part of the Bloch 
functions is $2\pi/k_\mu$, incommensurate with the lattice period. 

The oscillatory behavior of the single donor wave functions in Si, 
illustrated in Fig.~\ref{charge},  is well established experimentally \cite{Feher} 
and theoretically \cite{Koiller2004,overhof04}. 
This behavior does not bring significant consequences for conventional 
applications in Si-based devices (n-doped Si). 
A recent study of the single-qubit operations ($A$-gate) in the Kane QC shows that the 
$A$-gate operations do not present additional complications 
due to the Si band structure interference effects \cite{martins03}.

\section{Donor pair exchange coupling}

The HL approximation is a reliable scheme for the well-separated 
donor pair problem (interdonor distance much larger than the donor 
Bohr radii) \cite{slater}. Within  HL, the lowest energy singlet
and triplet wavefunctions for two electrons bound to a donor pair 
at sites $\mathbf{R}_A$ and $\mathbf{R}_B$, 
are written as properly symmetrized combinations of $\psi_{\mathbf{R}_A}$ 
and $\psi_{\mathbf{R}_B}$ [as defined in Eq.(\ref{eq:sim})]
\begin{eqnarray}
\Psi^s_t({\mathbf r}_1,{\mathbf r}_2) 
 &=& \frac{1}{\sqrt{2(1\pm S^2)}}
[ \psi_{{\bf R}_A}({\bf r}_1) \psi_{{\bf R}_B}({\bf r}_2)
\nonumber
\\
   &\pm& \psi_{{\bf R}_B}({\bf r}_1) \psi_{{\bf R}_A}({\bf r}_2)],
\label{eq:hl}
\end{eqnarray}
where $S$ is the overlap integral and the upper (lower) sign corresponds to
the singlet (triplet) state.  
The energy expectation values for these states, 
$E^s_t = \langle\Psi^s_t|{\cal H}|\Psi^s_t\rangle$, 
gives the exchange splitting through their difference, $J=E_t-E_s$.
We have previously derived the expression for the donor electron
exchange splitting \cite{KHD2,Koiller2004}, which we reproduce here:
\begin{eqnarray} 
J({\bf R}) = \frac{1}{36}\sum_{\mu, \nu} {\cal J}_{\mu \nu}
({\bf R}) \cos ({\bf k}_{\mu}-{\bf k}_{\nu})\cdot {\bf R}\,,
\label{eq:exch}
\end{eqnarray}
where $\mathbf{R} = \mathbf{R}_A - \mathbf{R}_B$ is the interdonor position vector 
and ${\cal J}_{\mu \nu} ({\bf R})$ are kernels determined by the envelopes
and are slowly varying \cite{KHD1,KHD2}.  
Note that equation~(\ref{eq:exch}) does not involve any oscillatory contribution from 
$u_{\mu}({\bf r})$, the periodic part of the Bloch
functions \cite{wellard03,Koiller2004}. The physical reason for that is clear from 
(\ref{eq:sim}): While the plane-wave phases of the Bloch functions are pinned to the  
donor sites, leading to the cossine factors in (\ref{eq:exch}), the periodic functions 
$u_\mu$ are pinned to the lattice, regardless of the donor location. 

\begin{figure}
\resizebox{75mm}{!}{\includegraphics[width=4.1in]
{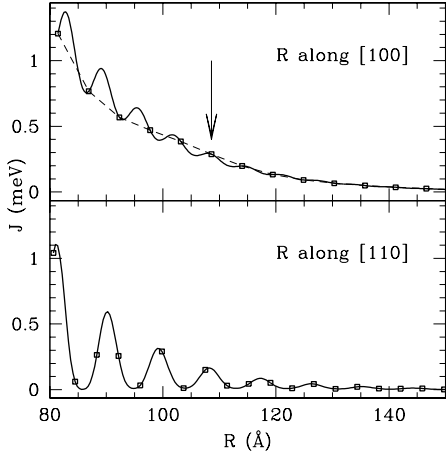}}
\caption
{Exchange coupling between two phosphorus donors in Si along the indicated
directions in the diamond structure. Values appropriate for impurities at
substitutional sites are given by the squares.
The dashed line in the $R \| [100]$ frame is a guide to the eye, indicating that
the oscillatory 
behavior may be ignored for donors positioned exactly along this axis.
}
\label{fig:oscillate}
\end{figure}

The exchange energy calculated from Eq.~(\ref{eq:exch}) for a pair of donors 
as a function of their relative position along the [100] and [110] crystal
axis is given in Fig.~\ref{fig:oscillate}.
This figure vividly illustrates both the anisotropic and the oscillatory
behavior of $J({\bf R})$, 
which is well established from previous
studies \cite{Andres,KHD1,KHD2,wellard03}.
It is interesting to note that for substitutional donors with 
interdonor position vectors exactly aligned with the 
[100] crystal axis, the oscillatory behavior may be ignored in practice, as
indicated by the dashed line in the figure.
This behavior is qualitatively similar to the exchange versus donor separation 
dependence assumed in Kane's proposal \cite{Kane}, where the Herring and Flicker 
expression \cite{herring64}, originally derived for H atoms,  was adapted for donors in Si.
Therefore one might expect that reliable exchange gates operation would be
possible if donors are exactly aligned along the [100] crystal axis.  

\section{Nanofabrication aspects}

\begin{figure}
\resizebox{75mm}{!}{\includegraphics[width=4.1in]
{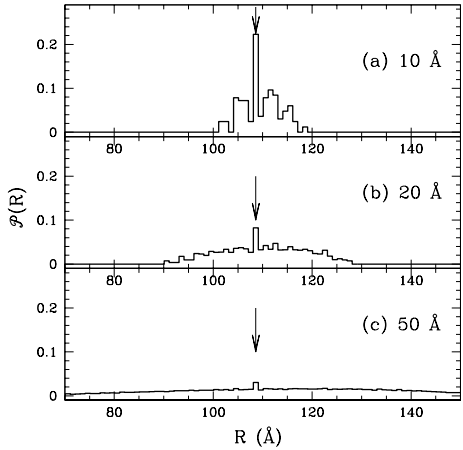}}
\caption{
Interdonor distance distributions for a {\it target} relative position of 20 
lattice parameters along [100] (see arrows). The first donor is fixed
and the second one ``visits'' all of the Si substitutional lattice 
sites within a sphere centered at the {\it target} position, 
with uncertainty radii (a)10 \AA, (b)20 \AA~and (c)50 \AA. 
}
\label{fig:histo}
\end{figure}

Aiming at the fabrication of a P donor array accurately positioned along the
[100] axis, and given the current degree of control in substitutional P positioning in 
Si of a few nm \cite{Obrien,encapsulation,implant,schenkel03}, we investigate 
the consequences of interdonor positioning uncertainties in the 
values of the corresponding pairwise exchange coupling.  
We define the {\it target} interdonor position ${\bf R}_t$ along [100],  
with an arbitrarily chosen length of 20 lattice parameters 
$(\sim 108.6$ \AA) indicated by the arrow in 
Fig.~\ref{fig:oscillate}. 
The distributions for the interdonor distances $R = |{\bf R}_A-{\bf R}_B|$  
when ${\bf R}_A$ is fixed and ${\bf R}_B$ ``visits'' all of the diamond
lattice sites within a sphere centered at the {\it target} position 
are given in Fig.~\ref{fig:histo}. Different frames give results for different 
uncertainty radii, and, as expected, increasing the uncertainty radius results in a broader
distribution around the {\it target} distance. Note that the 
geometry of the lattice implies that the distribution is always centered and 
peaked around $R_t$, as indicated by the arrows.  The additional peaks in the
distribution reveal the discrete nature of the Si lattice.

\begin{figure}
\resizebox{75mm}{!}{\includegraphics[width=4in]
{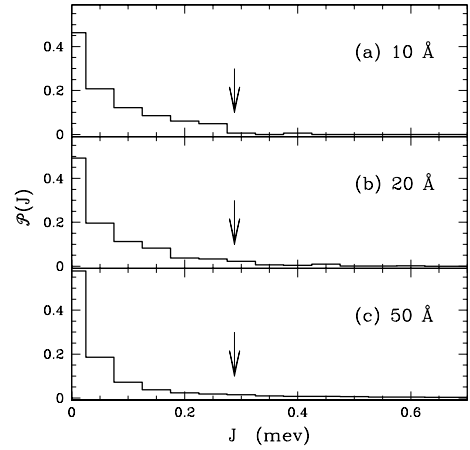}}
\caption{
Exchange distributions for the same relative position {\it ensembles} in
Fig.~\ref{fig:histo}.
The arrow indicates the {\it target} situation. Contrary to the distance
distributions, 
the exchange distributions are not centered or peaked around the target value.
}
\label{fig:histoj}
\end{figure}

The respective distributions of exchange coupling between the same donor pairs 
in each {\it ensemble} is presented in Fig.~\ref{fig:histoj}, where the arrows
give the exchange value at the {\it target} relative position: $J({\bf R}_t) \sim 0.29$
meV.
The results here are qualitatively different from the distance distributions
in 
Fig.~\ref{fig:histo}, since they are neither centered nor peaked at the {\it
target} 
exchange value. Even for the smallest uncertainty radius of 1 nm in (a), 
the exchange distribution is peaked around $J\sim 0$, 
bearing no semblance to the inter-donor distance distributions.
Increasing the uncertainty radius leads to a wider range of exchange values, 
with a more pronounced peak around the lowest $J$ values.

From the perspective of current QC fabrication efforts, 
$\sim 1$ nm accuracy in single P atom positioning has been 
recently demonstrated \cite{encapsulation}, representing
a major step towards the goal of obtaining a regular donor array 
embedded in single crystal Si. 
Distances and exchange coupling distributions consistent with such accuracy
are 
presented in Figs.\ref{fig:histo}(a) and \ref{fig:histoj}(a) respectively.
The present calculations indicate that such 
deviations in the relative position of donor pairs with respect to
perfectly aligned substitutional sites along [100] lead to order-of-magnitude
changes in the exchange coupling, favoring $J\sim 0$ values.  
Severe limitations in controlling $J$ would come from ``hops'' 
into different substitutional lattice sites. 
Therefore, precisely controlling exchange gates in Si remains an open
challenge.

%

\begin{theacknowledgments}
This work was partially supported by CNPq and by Instituto do Mil\^enio de
Nanoci\^encias in Brazil, by ARDA and LPS at the University of Maryland 
and by ARDA and ARO at the University at Buffalo. 
\end{theacknowledgments}

\bibliographystyle{aipproc}

\bibliography{float1}


\end{document}